\documentclass[sigconf, noacm]{acmart}

\graphicspath{{./images/}} 
\usepackage{soul}
\usepackage{graphicx}
\usepackage{caption}
\usepackage{multirow,multicol}
\begin{document}

\title{Evaluating the Performance of the DeepSeek Model in Confidential Computing Environment}


\author{Ben Dong}
\affiliation{%
  \institution{University of California, Merced}
  \city{Merced}
  \country{USA}}
\email{cdong12@ucmerced.edu}

\author{Qian Wang}
\affiliation{%
  \institution{University of California, Merced}
  \city{Merced}
  \country{USA}
}
\email{qianwang@ucmerced.edu}



\settopmatter{printacmref=false} 
\renewcommand\footnotetextcopyrightpermission[1]{} 
\acmConference[]{}{ }{ }

\begin{abstract}
The increasing adoption of Large Language Models (LLMs) in cloud environments raises critical security concerns, particularly regarding model confidentiality and data privacy. Confidential computing, enabled by Trusted Execution Environments (TEEs), offers a promising solution to mitigate these risks. However, existing TEE implementations, primarily CPU-based, struggle to efficiently support the resource-intensive nature of LLM inference and training. In this work, we present the first evaluation of the DeepSeek model within a TEE-enabled confidential computing environment, specifically utilizing Intel Trust Domain Extensions (TDX). Our study benchmarks DeepSeek’s performance across CPU-only, CPU-GPU hybrid, and TEE-based implementations. For smaller parameter sets, such as DeepSeek-R1-1.5B, the TDX implementation outperforms the CPU version in executing computations within a secure environment. It highlights the potential for efficiently deploying LLM models on resource-constrained systems while ensuring security. The overall GPU-to-CPU performance ratio averages 12 across different model sizes, with smaller models exhibiting a lower ratio. Additionally, we provide foundational insights and guidance on optimizing CPU-GPU confidential computing solutions for scalable and secure AI deployments. Our findings contribute to the advancement of privacy-preserving AI, paving the way for efficient and secure LLM inference in confidential computing environments.
\end{abstract}




\maketitle

\section{Introduction}
Cloud computing provides scalability and flexibility for computational tasks, particularly in machine learning applications. However, deploying ML models and processing sensitive data in cloud infrastructure introduces significant security risks. For example, unauthorized access to confidential user data can compromise data security within the cloud environment \cite{wallace2020concealed,tramer2016stealing}. This is especially critical in sectors like healthcare, where patient records could be exposed, or finance, where unauthorized access to banking transactions could lead to fraud and identity theft. Additionally, state-of-the-art machine learning models, such as large language models, require substantial financial investment for training and development. When deployed in cloud environments, they become vulnerable to theft in untrusted cloud infrastructure. 

In this work, we consider a threat model where both the user data and the machine learning model are at risk of leakage when processed in the shared cloud environment. Sensitive information, including personal photos, health records, and proprietary datasets, is at risk of exposure through reverse data inference attacks. Also, the integrity of ML models transmitted and executed in the cloud is susceptible to adversarial manipulations, including model inversion and poisoning attacks \cite{carlini2020extracting}. Protecting both data and model confidentiality is therefore critical.

Confidential computing, particularly through Trusted Execution Environments, provides a promising solution to secure both data and models. TEE provides a secure enclave for computation, protecting both data and models while mitigating risks in cloud computing environments. Prior research has leveraged TEEs, such as Intel’s Software Guard Extensions (SGX) \cite{intel_sgx_whitepaper}, to secure ML workloads by isolating sensitive computations within secure enclaves \cite{narra2019privacy}. While SGX offers strong protection, its limited memory capacity (approximately 1 GB) and complex interface pose challenges for large-scale ML applications \cite{shen2022soter,sun2023shadownet}. To overcome these limitations, Intel’s Trust Domain Extensions (TDX) \cite{intel_tdx_whitepaper} introduces secure Virtual Machines (VMs) that can process larger ML models while maintaining robust security guarantees.

Even though TDX significantly expands memory capacity compared to its predecessors, deploying the advanced large language models, such as GPT \cite{brown2020language} and Gemini \cite{deepmind2023gemini}, in a confidential computing environment presents additional challenges. First, these models are relatively large, typically starting from at least 1 billion parameters and reaching over 100 billion or more. Even hosting these models locally poses a significant challenge due to their high memory requirements.
Additionally, these models demand extensive computational resources, including high-performance GPUs for both training and inference. Ensuring security while maintaining efficiency in TEE-based environments requires careful optimization of both hardware and software components. 

Among these LLMs, DeepSeek is an advanced AI-driven model optimized for efficient resource utilization and high-performance computing \cite{bi2024deepseek}. Its main difference from other models lies in its ability to intelligently allocate resources using cutting-edge algorithms to ensure superior computational efficiency. It also enhances scalability, particularly in secure environments like TDX. For example, the distilled version of the DeepSeek model offers significant advantages by reducing model size and memory footprint, enabling intelligent resource allocation for optimal performance, even in constrained environments \cite{guo2024deepseek}. This is especially valuable in secure environments like TDX, where both efficiency and security are critical.
Exploring DeepSeek in TDX enables us to leverage its capabilities for secure and high-performance computing, facilitating faster data processing. This integration not only protects models from various security threats but also drives innovation in secure computing environments. Moreover, it provides insights into effectively addressing the security vs. performance trade-off, ensuring an optimal balance between protection and efficiency.

\subsection*{Contribution}

The main contributions of this paper are summarized as follows,
\begin{itemize}
    \item This paper presents the first evaluation of the DeepSeek model's performance within a TEE. By deploying DeepSeek in a secure enclave, we analyze how TEEs impact computational efficiency, resource utilization, and the inference performance. Our study provides a baseline understanding of DeepSeek’s behavior in confidential computing environments, offering critical insights into its feasibility for secure AI inference.
    \item We conduct a comparative analysis of DeepSeek’s performance across TEE-based, CPU-only, and CPU-GPU implementations. This benchmarking allows us to identify key performance bottlenecks and highlight trade-offs between security and computational efficiency. Our findings also help confidential computing technology vendors in optimizing CPU-GPU integration and addressing scalability challenges in secure AI workloads.
    \item Beyond performance evaluation, this study provides foundational insights for the broader research community on adopting CPU-GPU confidential computing solutions. By establishing a framework for evaluating large-scale AI models in TEEs, we contribute to the advancement of privacy-preserving AI technologies.
\end{itemize}



\section{Implementing LLM Models on TEE}

\subsection{Confidential Computing} Advanced LLMs are expensive to train and fine-tune due to their high computational and resource requirements. For instance, state-of-the-art models like ChatGPT-4 require thousands of GPUs (e.g., over 25,000 NVIDIA A100 GPUs) and months of training, with estimated costs exceeding \$100 million.  As a result, trained and fine-tuned models, particularly their weights, are highly sensitive intellectual property (IP) that must be safeguarded against unauthorized access and misuse. Additionally, during inference, user inputs may contain confidential or personally identifiable information (PII), which must be protected from potential security threats and malicious actors to ensure data privacy and integrity.

\begin{figure}[htb]
    \centering
    \captionsetup{font=small} 
    \includegraphics[width=1.1\linewidth]{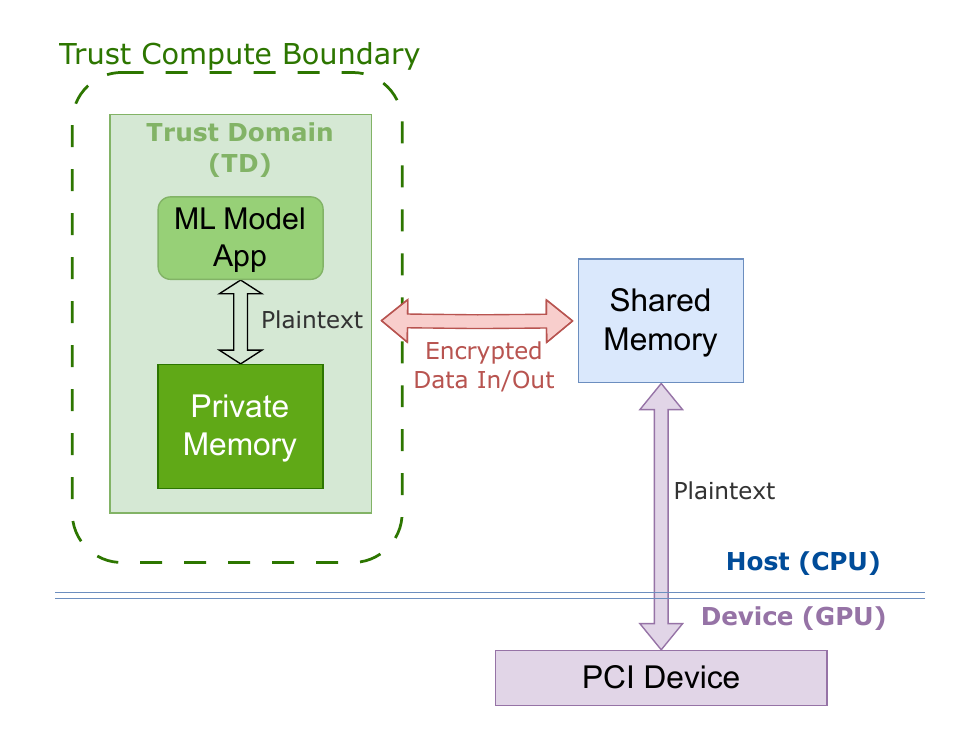}
    \caption{Trust Environment Settings with Data I/O to Shared memory and GPU.}
    \label{fig:tdx}
\end{figure}

Confidential computing offers a robust solution by ensuring that all data and workloads are executed within a secure enclave, where input-output communications are encrypted, shown as in Figure \ref{fig:tdx}. The Virtual Machine container, secured within a trusted enclave, ensures a confidential execution environment with exclusive access to allocated private memory. All plaintext computations occur securely within this isolated domain. Data exchanges between the enclave and shared memory undergo encryption and decryption. Additionally, the system interfaces with a PCI-connected device to facilitate collaborative processing between the CPU and GPU to optimize workload distribution for LLM computations.
This prevents unauthorized access and mitigates attacks to extract private data. However, current confidential computing solutions primarily rely on CPU-based TEEs with limited resources, restricting their ability to support large models like LLMs. This constraint creates a trade-off between security and performance, as secure execution environments struggle to handle the computational demands of large-scale AI models efficiently. 

Here, we evaluate the performance trade-offs between CPU-based confidential computing (CPU-TDX), standard CPU execution (CPU only), and GPU-accelerated platforms (CPU-GPU). Our study provides valuable insights for selecting the optimal balance between security and computational efficiency in deploying confidential inference workloads. 


\subsection{LLM Models} 
For this evaluation, we employed DeepSeek R1, a state-of-the-art reasoning-oriented large language model, across three different parameter configurations: 1.5 billion (1.5B), 7 billion (7B), and 14 billion (14B). DeepSeek R1 is designed for efficient logical reasoning and problem-solving, making it well-suited for workloads where reasoning and logic are critical. Running localized LLM inference within a confidential computing environment provides enhanced security guarantees by ensuring that both model weights and user input remain protected from unauthorized access or leakage. This is particularly important for enterprises and organizations handling sensitive data, as it mitigates risks associated with model inversion attacks, data breaches, and side-channel attacks. By leveraging Intel TDX, we analyze the trade-offs between security and computational performance when deploying DeepSeek R1 within a trusted execution environment, offering insights into the feasibility of secure and efficient AI inference.

\section{Experimental Results}
\subsection{Evaluation Platform}
The evaluation was conducted using a high-performance host machine equipped with two Intel Xeon Gold 6530 CPUs, each featuring 32 cores and 512 GB of DDR5 memory operating at 4800 MHz. The testing environment was deployed within an Intel Trust Domain Extensions virtual machine, configured according to Canonical Ubuntu’s official installation guide. The host system's BIOS settings were adjusted to enable TDX support, ensuring the correct hardware configuration for secure execution. To benchmark the performance of LLM inference, Ollama was executed directly within the TDX VM. A comparative evaluation was performed using Docker containers to simulate different execution environments. Specifically, a container was launched from Ollama’s official image with CPU and memory constraints simulating the TDX VM configuration. Additionally, a separate container was created with GPU access using NVIDIA’s container toolkit to assess the performance differential between CPU-based execution and GPU acceleration. All performance data was gathered using Ollama’s built-in logging mechanisms, ensuring consistency across test conditions. This setup enables a comprehensive analysis of inference efficiency in confidential computing environments while highlighting the trade-offs between security and computational performance.
\hspace{-0.5cm}
\begin{figure}[htb]
\hspace{-0.5cm}
    \centering
    \hspace{-0.5cm}
    \captionsetup{font=small} 
    \hspace{-0.5cm}
    \includegraphics[width=1.1\linewidth]{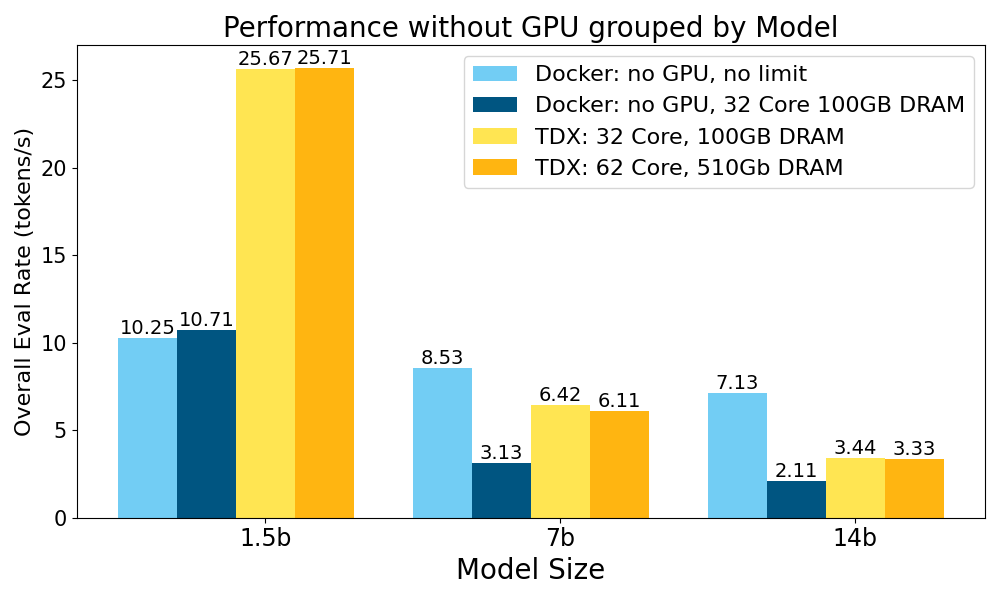}
    \hspace{-0.5cm}
    \caption{Model Performance on CPU with and w/o TDX.}
    \hspace{-0.5cm}
    \label{fig:performance}
    \hspace{-0.5cm}
\end{figure}

\subsection{Inference Workload} LLM inference in a TEE involves executing a trained machine-learning model to generate outputs based on user input prompts while ensuring data confidentiality and model integrity. In this benchmark, we evaluate different configurations where either a CPU or a CPU-GPU combination handles the workload. Typically, in standard LLM inference, the CPU is responsible for data loading, preprocessing, and postprocessing, while the GPU accelerates core computations. However, in confidential computing mode, the enclave-CPU takes on additional responsibilities for secure data processing to maintain model integrity and data privacy. The inference workflow begins with loading the trained machine learning model into the memory allocated to the Trusted Domain (TD). Once inside the TD, computations occur in an isolated environment without direct communication with the untrusted external environment, ensuring that sensitive data and model parameters remain protected from potential threats. This evaluation provides insights into the performance impact of running LLM inference within a TEE and highlights the trade-offs between security and computational efficiency across different hardware configurations.

\begin{table}[h]
\captionsetup{font=small} 
\centering
\resizebox{\columnwidth}{!}{ 
\begin{tabular}{c||c|c|c|c|c}
\hline
\multirow{2}{*}{\textbf{Model Size}} & \multicolumn{3}{c|}{\textbf{Performance (tokens/s)}} & \multicolumn{2}{c}{\textbf{Comp. Ratio}} \\ \cline{2-6}
 & \textbf{GPU-CPU} & \textbf{CPU only} & \textbf{TDX} & \textbf{$\frac{GPU}{TDX}$} & \textbf{$\frac{CPU}{TDX}$} \\ \hline
1.5B & 202.88  & 10.25  & 25.67 & 7.9 & 0.4 \\ \hline
7B    & 117.02 & 8.53  & 6.42  & 18.2 & 1.3 \\ \hline
14B & 69.14 & 7.13 & 3.44 & 9.7 & 2 \\ \hline
\end{tabular}
}
\caption{Comparison of Performance by Model With GPU (Tokens/s): GPU-CPU represents inference running directly on the host machine with GPU accelerations enabled. CPU-only refers to running on the host machine without GPU acceleration enabled. TDX runs within a TD container and restricted to 32 CPU cores and 100GB of DRAM.}
\label{tab:model_performance_gpu}
\end{table}

\subsection{Result Analysis} The results from both the figure and the table highlight the performance differences between Docker-based CPU execution and Intel TDX, suggesting that TDX's optimizations for secure computing may also enhance CPU performance in certain scenarios. In CPU-only configurations (as shown in Figure \ref{fig:performance}), the TDX environment with 62 cores and 510GB of DRAM achieves the highest evaluation rate for the smallest model (1.5B), reaching approximately 25.71 tokens/s—more than twice the performance observed in Docker-based CPU tests. This significant improvement could be attributed to TDX's optimized CPU execution, minimizing software-induced inefficiencies. However, as the size of the model increases, the performance gap narrows, with both TDX and Docker CPU-based environments showing significant slowdowns for the 14B model, where evaluation rates are limited to around 3.33 tokens/s. This suggests that while TDX can leverage high-core configurations effectively for smaller models, larger models demand higher memory bandwidth and computational resources, limiting the advantage of TDX's CPU optimizations.

In contrast, Table \ref{tab:model_performance_gpu} shows that GPU acceleration drastically improves inference speed across all configurations, far surpassing CPU-only performance. Even with resource constraints, GPU-based inference reaches over 202 tokens/s for the 1.5B model, significantly outperforming any CPU configuration. The 7B and 14B models also maintain high performance, achieving approximately 117 and 69 tokens/s, respectively. These results emphasize the critical role of GPU acceleration in the efficient deployment of large LLMs. However, current TDX implementations do not fully support GPU utilization within the secure enclave, presenting a key limitation in balancing security and performance. Future research should focus on integrating GPU acceleration within TDX while preserving its security guarantees, enabling confidential AI inference to achieve both high security and computational efficiency.

\section{Discussion}
Our evaluation of the DeepSeek model in a TEE underscores the complexities and trade-offs involved in the deployment of LLMs securely. Although TEEs, such as Intel-TDX, offer robust protections against unauthorized access, the reliance on CPU-based enclaves introduces performance limitations, particularly for compute-intensive workloads with larger size of models. GPU acceleration is essential for efficient model execution, yet current confidential computing frameworks do not fully support GPU-based processing within secure enclaves. This limitation results in a significant trade-off between maintaining security and achieving high-performance inference.

Furthermore, our results highlight key considerations for optimizing confidential AI deployments. The computational overhead of maintaining security, such as memory encryption and secure data handling within the enclave, leads to increased inference latency. However, as confidential computing technologies evolve, emerging solutions like secure GPU virtualization and hybrid execution models may help mitigate these overheads. Future research should focus on refining these technologies to enable efficient, secure AI processing without sacrificing performance.

\section{Conclusion}

This paper presents the first performance evaluation of the DeepSeek model in a confidential computing environment, comparing CPU-based TEEs, standard CPU execution, and GPU-accelerated platforms. Our findings provide critical insights into the feasibility of running LLM inference securely, highlighting the need for balancing security and computational efficiency. While CPU-based TEE ensures strong model integrity and data confidentiality, its performance constraints necessitate further advancements in confidential computing frameworks, particularly in enabling GPU acceleration. To sum up, this work serves as a foundation for future studies, guiding the development of scalable and efficient confidential computing solutions for AI applications.


\bibliographystyle{ACM-Reference-Format}
\bibliography{references.bib}


\begin{thebibliography}{12}


\ifx \showCODEN    \undefined \def \showCODEN     #1{\unskip}     \fi
\ifx \showDOI      \undefined \def \showDOI       #1{#1}\fi
\ifx \showISBNx    \undefined \def \showISBNx     #1{\unskip}     \fi
\ifx \showISBNxiii \undefined \def \showISBNxiii  #1{\unskip}     \fi
\ifx \showISSN     \undefined \def \showISSN      #1{\unskip}     \fi
\ifx \showLCCN     \undefined \def \showLCCN      #1{\unskip}     \fi
\ifx \shownote     \undefined \def \shownote      #1{#1}          \fi
\ifx \showarticletitle \undefined \def \showarticletitle #1{#1}   \fi
\ifx \showURL      \undefined \def \showURL       {\relax}        \fi
\providecommand\bibfield[2]{#2}
\providecommand\bibinfo[2]{#2}
\providecommand\natexlab[1]{#1}
\providecommand\showeprint[2][]{arXiv:#2}

\bibitem[Bi et~al\mbox{.}(2024)]%
        {bi2024deepseek}
\bibfield{author}{\bibinfo{person}{Xiao Bi}, \bibinfo{person}{Deli Chen}, \bibinfo{person}{Guanting Chen}, \bibinfo{person}{Shanhuang Chen}, \bibinfo{person}{Damai Dai}, \bibinfo{person}{Chengqi Deng}, \bibinfo{person}{Honghui Ding}, \bibinfo{person}{Kai Dong}, \bibinfo{person}{Qiushi Du}, \bibinfo{person}{Zhe Fu}, {et~al\mbox{.}}} \bibinfo{year}{2024}\natexlab{}.
\newblock \showarticletitle{Deepseek llm: Scaling open-source language models with longtermism}.
\newblock \bibinfo{journal}{\emph{arXiv preprint arXiv:2401.02954}} (\bibinfo{year}{2024}).
\newblock


\bibitem[Brown et~al\mbox{.}(2020)]%
        {brown2020language}
\bibfield{author}{\bibinfo{person}{Tom Brown}, \bibinfo{person}{Benjamin Mann}, \bibinfo{person}{Nick Ryder}, \bibinfo{person}{Melanie Subbiah}, \bibinfo{person}{Jared Kaplan}, \bibinfo{person}{Prafulla Dhariwal}, \bibinfo{person}{Arvind Neelakantan}, \bibinfo{person}{Pranav Shyam}, \bibinfo{person}{Girish Sastry}, \bibinfo{person}{Amanda Askell}, {et~al\mbox{.}}} \bibinfo{year}{2020}\natexlab{}.
\newblock \showarticletitle{Language models are few-shot learners}.
\newblock \bibinfo{journal}{\emph{Advances in Neural Information Processing Systems (NeurIPS)}}  \bibinfo{volume}{33} (\bibinfo{year}{2020}), \bibinfo{pages}{1877--1901}.
\newblock
\urldef\tempurl%
\url{https://arxiv.org/abs/2005.14165}
\showURL{%
\tempurl}


\bibitem[Carlini et~al\mbox{.}(2020)]%
        {carlini2020extracting}
\bibfield{author}{\bibinfo{person}{Nicholas Carlini}, \bibinfo{person}{Florian Tramer}, \bibinfo{person}{Eric Wallace}, \bibinfo{person}{Matthew Jagielski}, \bibinfo{person}{Ariel Herbert-Voss}, \bibinfo{person}{Katherine Lee}, \bibinfo{person}{Adam Roberts}, \bibinfo{person}{Tom Brown}, \bibinfo{person}{Dawn Song}, \bibinfo{person}{Ulfar Erlingsson}, \bibinfo{person}{Alina Oprea}, {and} \bibinfo{person}{Colin Raffel}.} \bibinfo{year}{2020}\natexlab{}.
\newblock \showarticletitle{Extracting Training Data from Large Language Models}.
\newblock \bibinfo{journal}{\emph{arXiv preprint arXiv:2012.07805}} (\bibinfo{year}{2020}).
\newblock


\bibitem[DeepMind(2023)]%
        {deepmind2023gemini}
\bibfield{author}{\bibinfo{person}{Google DeepMind}.} \bibinfo{year}{2023}\natexlab{}.
\newblock \bibinfo{title}{Introducing Gemini: Our most capable AI model}.
\newblock
\newblock
\urldef\tempurl%
\url{https://www.deepmind.com/blog/introducing-gemini}
\showURL{%
\tempurl}


\bibitem[Guo et~al\mbox{.}(2024)]%
        {guo2024deepseek}
\bibfield{author}{\bibinfo{person}{Daya Guo}, \bibinfo{person}{Qihao Zhu}, \bibinfo{person}{Dejian Yang}, \bibinfo{person}{Zhenda Xie}, \bibinfo{person}{Kai Dong}, \bibinfo{person}{Wentao Zhang}, \bibinfo{person}{Guanting Chen}, \bibinfo{person}{Xiao Bi}, \bibinfo{person}{Yu Wu}, \bibinfo{person}{YK Li}, {et~al\mbox{.}}} \bibinfo{year}{2024}\natexlab{}.
\newblock \showarticletitle{DeepSeek-Coder: When the Large Language Model Meets Programming--The Rise of Code Intelligence}.
\newblock \bibinfo{journal}{\emph{arXiv preprint arXiv:2401.14196}} (\bibinfo{year}{2024}).
\newblock


\bibitem[{Intel Corporation}(2019)]%
        {intel_sgx_whitepaper}
\bibfield{author}{\bibinfo{person}{{Intel Corporation}}.} \bibinfo{year}{2019}\natexlab{}.
\newblock \bibinfo{title}{Overview on Signing and Whitelisting for Intel® Software Guard Extensions (Intel® SGX) Enclaves}.
\newblock \bibinfo{howpublished}{\url{https://www.intel.com/content/dam/develop/external/us/en/documents/overview-signing-whitelisting-intel-sgx-enclaves.pdf}}.
\newblock


\bibitem[{Intel Corporation}(2022)]%
        {intel_tdx_whitepaper}
\bibfield{author}{\bibinfo{person}{{Intel Corporation}}.} \bibinfo{year}{2022}\natexlab{}.
\newblock \bibinfo{title}{Intel® Trust Domain Extensions}.
\newblock \bibinfo{howpublished}{\url{https://cdrdv2-public.intel.com/690419/TDX-Whitepaper-February2022.pdf}}.
\newblock


\bibitem[Narra et~al\mbox{.}(2019)]%
        {narra2019privacy}
\bibfield{author}{\bibinfo{person}{Krishna~Giri Narra}, \bibinfo{person}{Zhifeng Lin}, \bibinfo{person}{Yongqin Wang}, \bibinfo{person}{Keshav Balasubramaniam}, {and} \bibinfo{person}{Murali Annavaram}.} \bibinfo{year}{2019}\natexlab{}.
\newblock \showarticletitle{Privacy-preserving inference in machine learning services using trusted execution environments}.
\newblock \bibinfo{journal}{\emph{arXiv preprint arXiv:1912.03485}} (\bibinfo{year}{2019}).
\newblock


\bibitem[Shen et~al\mbox{.}(2022)]%
        {shen2022soter}
\bibfield{author}{\bibinfo{person}{Tianxiang Shen}, \bibinfo{person}{Ji Qi}, \bibinfo{person}{Jianyu Jiang}, \bibinfo{person}{Xian Wang}, \bibinfo{person}{Siyuan Wen}, \bibinfo{person}{Xusheng Chen}, \bibinfo{person}{Shixiong Zhao}, \bibinfo{person}{Sen Wang}, \bibinfo{person}{Li Chen}, \bibinfo{person}{Xiapu Luo}, {et~al\mbox{.}}} \bibinfo{year}{2022}\natexlab{}.
\newblock \showarticletitle{$\{$SOTER$\}$: Guarding Black-box Inference for General Neural Networks at the Edge}. In \bibinfo{booktitle}{\emph{2022 USENIX Annual Technical Conference (USENIX ATC 22)}}. \bibinfo{pages}{723--738}.
\newblock


\bibitem[Sun et~al\mbox{.}(2023)]%
        {sun2023shadownet}
\bibfield{author}{\bibinfo{person}{Zhichuang Sun}, \bibinfo{person}{Ruimin Sun}, \bibinfo{person}{Changming Liu}, \bibinfo{person}{Amrita~Roy Chowdhury}, \bibinfo{person}{Long Lu}, {and} \bibinfo{person}{Somesh Jha}.} \bibinfo{year}{2023}\natexlab{}.
\newblock \showarticletitle{Shadownet: A secure and efficient on-device model inference system for convolutional neural networks}. In \bibinfo{booktitle}{\emph{2023 IEEE Symposium on Security and Privacy (SP)}}. IEEE, \bibinfo{pages}{1596--1612}.
\newblock


\bibitem[Tram{\`e}r et~al\mbox{.}(2016)]%
        {tramer2016stealing}
\bibfield{author}{\bibinfo{person}{Florian Tram{\`e}r}, \bibinfo{person}{Fan Zhang}, \bibinfo{person}{Ari Juels}, \bibinfo{person}{Michael~K Reiter}, {and} \bibinfo{person}{Thomas Ristenpart}.} \bibinfo{year}{2016}\natexlab{}.
\newblock \showarticletitle{Stealing Machine Learning Models via Prediction APIs}. In \bibinfo{booktitle}{\emph{25th {USENIX} Security Symposium ({USENIX} Security 16)}}. \bibinfo{pages}{601--618}.
\newblock


\bibitem[Wallace et~al\mbox{.}(2020)]%
        {wallace2020concealed}
\bibfield{author}{\bibinfo{person}{Eric Wallace}, \bibinfo{person}{Shi Feng}, \bibinfo{person}{Nikhil Kandpal}, \bibinfo{person}{Sameer Singh}, {and} \bibinfo{person}{Matt Gardner}.} \bibinfo{year}{2020}\natexlab{}.
\newblock \showarticletitle{Concealed Data Poisoning Attacks on NLP Models}. In \bibinfo{booktitle}{\emph{Proceedings of the 2020 Conference on Empirical Methods in Natural Language Processing (EMNLP)}}. \bibinfo{pages}{139--149}.
\newblock


\end{thebibliography}

\end{document}